\documentstyle[12pt,epsfig,a41,amssymb]{article}
\begin{document}
\thispagestyle{empty}
\begin{flushright}
\large
DO-TH 97/22 \\ 
DTP/97/90 \\
October 1997
\end{flushright}
\vspace{1.2cm}

\renewcommand{\thefootnote}{\fnsymbol{footnote}}
\setcounter{footnote}{1}
\begin{center}
{\LARGE\bf Radiative Parton Model Analysis}

\vspace*{0.5cm}
{\LARGE\bf of Recent Polarized DIS Data}

\vspace*{1.8cm}
{\Large Marco Stratmann}\footnote[1]{Invited talk presented at the
workshop 'Deep Inelastic Scattering off Polarized Targets:
Theory Meets Experiment', DESY-Zeuthen, Germany, September 1-5, 1997}

\vspace*{1.5cm}
{\it Institut f\"{u}r Physik, Universit\"{a}t Dortmund, 
44221 Dortmund, Germany}\\

\vspace*{2mm}
{\it and}\\

\vspace*{2mm}
{\it Department of Physics, University of Durham, 
Durham DH1 3LE, England\hspace*{1mm}\footnote[2]{Present address}} 

\vspace*{2.cm}

\renewcommand{\thefootnote}{\arabic{footnote}}
\setcounter{footnote}{0}
{\large \bf Abstract} 
\end{center}

\noindent
An updated next-to-leading order (NLO) QCD analysis of the spin
asymmetries $A_1^N(x,Q^2)$ and parton distributions 
$\delta f(x,Q^2)$ in longitudinally polarized deep-inelastic
lepton-nucleon scattering is presented within the framework of the
radiative parton model taking into account recent experimental
results. The theoretical framework and the main features of the
radiative parton model analysis are briefly reviewed.
The small-$x$ behaviour of the polarized structure function 
$g_1^N(x,Q^2)$ as well as the shape of the polarized 
gluon distribution $\delta g(x,Q^2)$ are shown to be still hardly
constrained by present fixed target data.

\setcounter{page}{0}
\newpage
%
%
\renewcommand{\thefootnote}{\fnsymbol{footnote}}
\setcounter{footnote}{1}
\begin{center}
{\LARGE\bf Radiative Parton Model Analysis}

\vspace*{0.5cm}
{\LARGE\bf of Recent Polarized DIS Data}

\vspace{1cm}
{\Large Marco Stratmann}

\vspace*{0.7cm}
{\it Institut f\"{u}r Physik, Universit\"{a}t Dortmund, 
44221 Dortmund, Germany}\\

\vspace*{2mm}
{\it and}\\

\vspace*{2mm}
{\it Department of Physics, University of Durham, 
Durham DH1 3LE, England\hspace*{1mm}\footnote{Present address}} 

\vspace*{1.cm}

\renewcommand{\thefootnote}{\arabic{footnote}}
\setcounter{footnote}{0}
\end{center}

%
%
\begin{abstract}
\noindent
An updated next-to-leading order (NLO) QCD analysis of the spin
asymmetries $A_1^N(x,Q^2)$ and parton distributions 
$\delta f(x,Q^2)$ in longitudinally polarized deep-inelastic
lepton-nucleon scattering is presented within the framework of the
radiative parton model taking into account recent experimental
results. The theoretical framework and the main features of the
radiative parton model analysis are briefly reviewed.
The small-$x$ behaviour of the polarized structure function 
$g_1^N(x,Q^2)$ as well as the shape of the polarized 
gluon distribution $\delta g(x,Q^2)$ are shown to be still hardly
constrained by present fixed target data.
\end{abstract}

%
%
\section{Introduction}
\noindent
The past year has seen much progress in our knowledge about the
nucleons' spin structure due to new experimental results on the
spin asymmetry $A_1^N(x,Q^2)\simeq g_1^N(x,Q^2)/F_1^N(x,Q^2)$ in
deep-inelastic scattering with longitudinally polarized lepton beams
off nucleon targets $(N=p,\,n,\,d)$. In particular, previous sparse
and not very precise experimental information on the neutron
asymmetry $A_1^n$ \cite{e142} has been succeeded by
more accurate data by the E154 collaboration \cite{e154} where also
the kinematical coverage in $x$ and $Q^2$ was slightly extended.
Recently the HERMES group also has presented first results on
$A_1^n$ \cite{hermesn} and preliminary data for a
proton target $(A_1^p)$ have been reported on this workshop
\cite{hermesp}. SMC has released a new detailed analysis of $A_1^p$
\cite{smcnew} which does not indicate a rise of $g_1^p$ at small-$x$
anymore and, finally, first very accurate, but still preliminary
results from E155 on $g_1^p$ were also presented on this workshop 
\cite{e155}.

\renewcommand{\thefootnote}{\arabic{footnote}}
\setcounter{footnote}{0}
In view of these recent experimental developments it seems to be 
worthwhile to reanalyse \cite{grsvnew} these data\footnote{The preliminary 
E155 \cite{e155} and HERMES \cite{hermesp} proton data are not 
available yet and hence not included in our analysis so far.} in terms 
of polarized parton distributions
$\delta f$ in the framework of perturbative QCD. 
Preceding studies \cite{grsv,gs,abfr} have revealed that the detailed 
$x$-shape of the polarized gluon distribution $\delta g(x,Q^2)$
was only weakly constrained by the
data of that time even though a tendency towards a sizeable positive 
{\em{total}} gluon polarization, $\int_0^1 \delta 
g(x,Q^2=4\,{\mathrm{GeV}}^2) dx \gtrsim 1$, was found \cite{grsv,gs,abfr}.
It it thus interesting to study to what extent these old results are 
confirmed by the new data sets and whether one can further 
pin down the polarized gluon distribution.

In the following we will concentrate exclusively on calculations of
$A_1^N$ to NLO accuracy which
became possible only recently after the derivation of the required
spin-dependent two-loop anomalous dimensions \cite{anom1,anom2}.
A first such complete and consistent NLO study has been presented in 
\cite{grsv} (based on a corresponding LO analysis \cite{grvpol}), 
where the underlying concept has been the radiative 
generation of parton distributions from a valence-like structure 
at some low bound-state like resolution scale $\mu$. In the unpolarized case
this had previously led \cite{grvold,grv94}, e.g., to the successful 
prediction of the small-$x$ rise of the proton structure function $F_2^p$ as
later on observed at HERA. Other NLO analyses of polarized DIS data 
can be found in \cite{gs,abfr}.

In the next section the basic theoretical framework for polarized DIS
beyond the leading order is briefly discussed hereby defining our
notations, and the main features of the radiative parton model
analysis \cite{grsv} are reviewed. Section 3 contains the presentation
and discussion of the new quantitative NLO QCD results and, finally, the main
findings are summarized in section 4.
%
\section{Theoretical Framework}
%
\noindent
Measurements of polarized deep inelastic lepton nucleon scattering yield
direct information [1-6,16] only on the spin-asymmetry
\begin{equation}
\label{eq:a1}
A_1^N (x,Q^2) \simeq \frac{g_1^N (x,Q^2)}{F_1^N (x,Q^2)}
=\frac{g_1^N(x,Q^2)}{F_2^N (x,Q^2)/ \left[ 2x(1+R^N(x,Q^2)) \right] } \:\:\: ,
\end{equation}
where $N=p,n,d$ (in the latter case we have used
$g_1^d=\frac{1}{2} (g_1^p+g_1^n) (1-\frac{3}{2} \omega_D)$ with
$\omega_D=0.058$) and $R \equiv F_L/2xF_1 =(F_2-2 x F_1)/2xF_1$. 
In NLO, the polarized structure function $g_1^N(x,Q^2)$ in (\ref{eq:a1})
is related to the spin-dependent quark and gluon distributions 
($\delta f^N$) in the following way:
\begin{equation}
\nonumber
g_1^N(x,Q^2) = \frac{1}{2} \sum_q e_q^2 \Bigg\{ \delta q^N(x,Q^2)+
\delta \bar{q}^N(x,Q^2)+
\frac{\alpha_s(Q^2)}{2\pi} \left[ \delta C_q \otimes \left( \delta q^N+
\delta \bar{q}^N\right) + \delta C_g \otimes \delta g\right]\!\! \Bigg\}
\label{eq:g1}
\end{equation}
with the convolutions ($\otimes$) being defined as usual.
The NLO pieces entering (\ref{eq:g1}), i.e., $\delta f^N$, $\delta C_q$, 
$\delta C_g$, depend on the factorization convention (scheme) adopted. 
Since the two-loop anomalous dimensions of \cite{anom1,anom2} refer to the
conventional $\overline{\mathrm{MS}}$ dimensional regularization
prescription we prefer to work also in this scheme.
The appropriate spin-dependent $\overline{\mathrm{MS}}$
Wilson coefficients $\delta C_q$ and
$\delta C_g$ can be found, e.g., in \cite{anom1,grsv}.
It should be recalled that the first moment
$\Gamma_1^N(Q^2)\equiv \int_0^1 g_1^N(x,Q^2)dx$ of eq.\ (\ref{eq:g1}) is
in the $\overline{\mathrm{MS}}$ scheme simply given by
\begin{equation}
\label{eq:gammag1}
\Gamma_1^N(Q^2)=\frac{1}{2} \sum_q e_q^2
\left( 1-\frac{\alpha_s(Q^2)}{\pi}
\right) \left[ \Delta q^N(Q^2)+\Delta \bar{q}^N(Q^2)\right]
\end{equation}
where we have introduced the first moments $\Delta f^N(Q^2)$ of the 
polarized distributions $\delta f^N(x,Q^2)$ by defining as usual 
\begin{equation}
\label{eq:firstmom}
\Delta f^N(Q^2) \equiv \int_0^1 \delta f^N(x,Q^2) dx
\end{equation}
where $f=u,\bar{u},d,\bar{d},s,\bar{s},$ and $g$ and have used (see, e.g., 
\cite{anom1,grsv}) $\int_0^1 \delta C_q(x) dx = -3 C_F/2$ and
$\int_0^1 \delta C_g(x) dx = 0$. Thus, the
total gluon helicity $\Delta g(Q^2)$ does not directly couple to
$\Gamma_1^N(Q^2)$ due to the vanishing of the integrated gluonic coefficient
function in the $\overline{\rm{MS}}$ factorization scheme. 
Other factorization schemes \cite{abfr} are of course also allowed 
but afford a proper scheme transformation (which cannot be
uniquely fixed) such that the physical 
quantity $g_1^N$ remains scheme independent up to ${\cal{O}}(\alpha_s)$.

The NLO $Q^2$-evolution of the polarized parton distributions $\delta f(x,Q^2)$
(henceforth we shall, as always, use the notation $\delta q^p\equiv \delta q$ 
and $q^p\equiv q$) governed by the two-loop anomalous dimensions  
\cite{anom1,anom2} is performed most conveniently in the Mellin-$n$ moment
space where the solutions of the evolution equations 
(see, e.g., refs.\cite{grvold,grsv}) can be obtained analytically, 
once the boundary conditions at some $Q^2=Q^2_0$, i.e., 
input densities $\delta f(x,Q_0^2)$ to be discussed below, are specified. 
Furthermore, in Mellin-$n$ space the
convolutions in (\ref{eq:g1}) reduce to simple products. Having obtained the
analytic NLO solutions for the moments of parton densities or the 
analogous $n$-space expression of (\ref{eq:g1}) the desired $x$-space
results for $\delta f(x,Q^2)$ or $g_1^N(x,Q^2)$ are then simply obtained by a
standard numerical Mellin inversion as described, e.g., in 
\cite{grvold}.

In our analysis only quarks with $m_q<\Lambda_{QCD}$, i.e., $u$, $d$, $s$,
will be treated as light partons in the evolution equations while the
charm contribution to $g_1^N$ (as well as to $F_1^N$) is calculated via the
appropriate massive $\gamma^* g\rightarrow c\bar{c}$ fusion process 
\cite{watson} although it turns out (see later) that in the kinematical 
region covered by present fixed target experiments [1-6,16] 
the charm contribution is extremely small and practically irrelevant.

To finally fix the NLO input parton densities $\delta f(x,Q_0^2)$ we perform 
fits only to the {\em{directly measured}} spin asymmetry $A_1^N(x,Q^2)$ in 
(\ref{eq:a1}), rather than to the derived $g_1^N(x,Q^2)$. 
The main reason for that is that in some 
older experimental analyses $g_1^N(x,Q^2)$ has been extracted under 
the assumption of the $Q^2$-independence of $A_1^N(x,Q^2)$, which is 
- although presently available data do not exhibit any significant
$Q^2$-dependence within the experimental errors - theoretically 
not warranted due
to the different $Q^2$-evolutions of the numerator and denominator in 
(\ref{eq:a1}). 

As already mentioned in the introduction, the other main 
ingredient of our NLO analysis \cite{grsv} is that we follow the radiative
(dynamical) concept \cite{grvold,grv94} by choosing the same low
input scale $Q_0^2=\mu^2=0.34\,\mathrm{GeV}^2$ and implementing the
fundamental positivity requirement
\begin{equation}
\label{eq:pos}
\left| \delta f(x,Q^2) \right | \le f(x,Q^2)
\end{equation}
down to $Q^2=\mu^2$.
The analysis affords some well established set of unpolarized NLO parton
distributions $f(x,Q^2)$ for calculating $F_1^N(x,Q^2)$ in 
(\ref{eq:a1}) and as reference distributions in (\ref{eq:pos}) which will 
be adopted from ref.\cite{grv94}.

In addition to (\ref{eq:pos}), the first moments $\Delta f(Q^2)$ of 
the NLO polarized parton distributions are taken to be subject to two very
different sets of theoretical constraints \cite{grvpol,grsv} 
related to two different views
concerning the flavor $SU(3)_f$ symmetry properties of hyperon
$\beta$-decays. One set ('standard' scenario) assumes an unbroken
$SU(3)_f$ symmetry between the relevant matrix elements leading
to the following sum rule constraints,
\vspace*{-1mm}
\begin{eqnarray}
\label{eq:stdsums1}
\Delta q_3 &=& \Delta u + \Delta \bar{u} - \Delta d - \Delta \bar{d} =
g_A = F + D = 1.2573 \pm 0.0028 \\
\label{eq:stdsums2}
\Delta q_8 &=& \Delta u + \Delta \bar{u} + \Delta d + \Delta \bar{d} -
2(\Delta s + \Delta \bar{s}) = 3F -D = 0.579 \pm 0.025
\vspace*{-1mm}
\end{eqnarray}
with the values of $g_A$ and $3F-D$ taken from \cite{fdvalues}.
It should be noted that the flavor
non-singlet combinations $\Delta q_{3,8}$ in (\ref{eq:stdsums1}) and 
(\ref{eq:stdsums2}) remain $Q^2$-independent also in NLO 
\cite{anom1,anom2,grsv}. 

As a plausible alternative to the full $SU(3)_f$ symmetry between charged
weak and neutral axial currents required for deriving the 'standard'
constraints (\ref{eq:stdsums1}) and (\ref{eq:stdsums2}), we consider 
a 'valence' scenario \cite{grvpol,lipkin} where this flavor symmetry is broken 
and which is based on the assumption \cite{lipkin} that the flavor changing
hyperon $\beta$-decay data fix only the total helicity of
{\em{valence}} quarks:
\vspace*{-1mm}
\begin{eqnarray}
\label{eq:valsums1}
\Delta u_V(\mu^2)- \Delta d_V(\mu^2) &=& g_A = F+D= 1.2573 \pm 0.0028 \\
\protect{\nopagebreak}
\label{valsums2}
\Delta u_V(\mu^2)+\Delta d_V(\mu^2) &=& 3F-D  = 0.579 \pm 0.025\;\;\;.
\end{eqnarray}
\vspace*{-2mm}
%
%
%
\begin{figure}[th]
\begin{center}
\vspace*{-1.5cm}
\epsfig{file=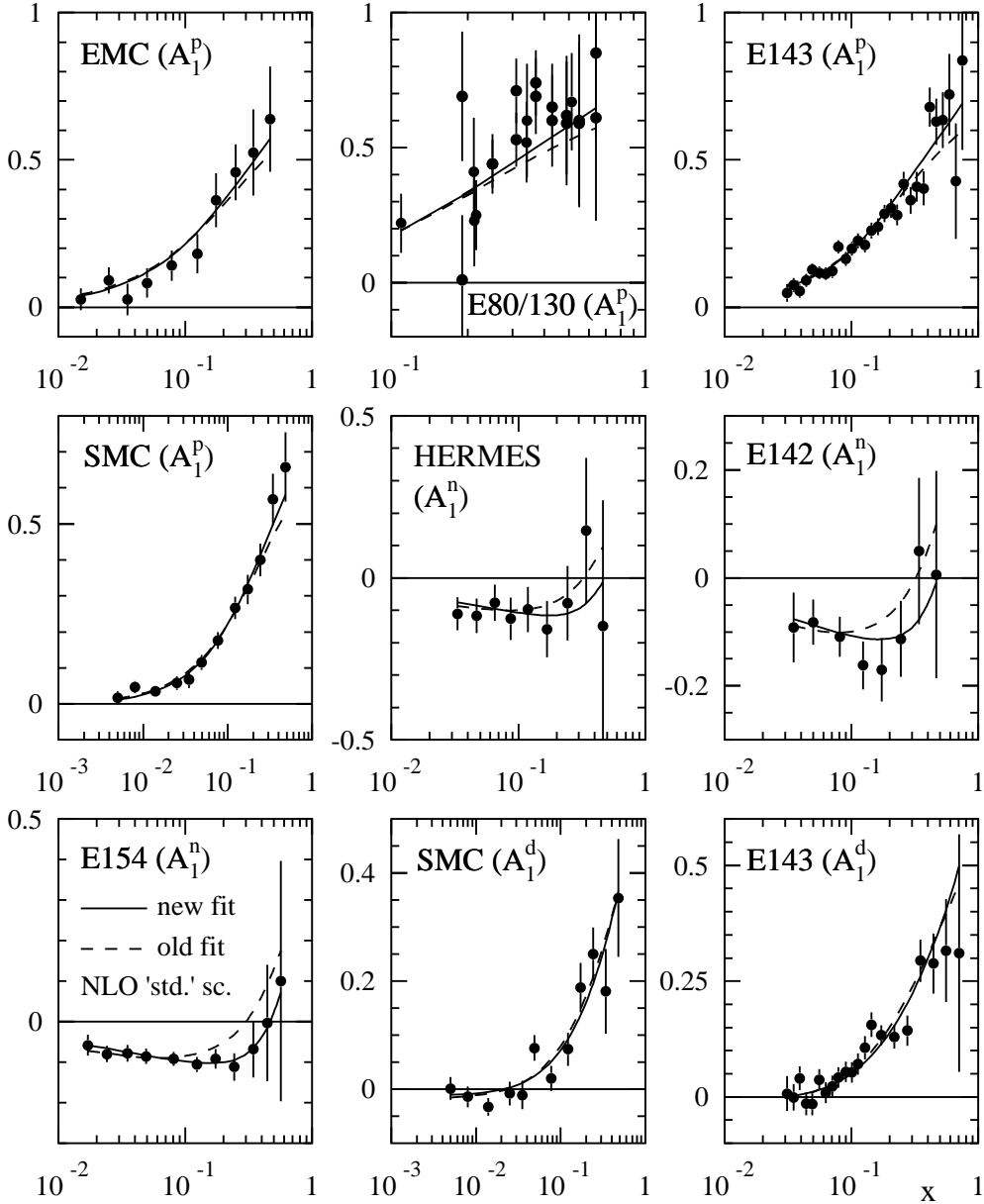,width=13.5cm}
\vspace*{-0.3cm}
\caption{\sf Comparison of our reanalysed NLO 
'standard' results for $A_1^N(x,Q^2)$
(solid lines) with all presently available data [1-3,5,16]. The $Q^2$ values
adopted here correspond to the different values quoted in 
[1-3,5,16]. Also shown are the results of our previous analysis \cite{grsv} 
(dashed lines).}
\vspace*{-0.5cm}
\end{center}
\end{figure}

The 'standard' scenario always requires a finite total
strange sea helicity of $\Delta s =\Delta \bar{s} \simeq -0.05$ in order
to account for the experimentally observed reduction of $\Gamma_1^p$ 
with respect to the Gourdin and Ellis and Jaffe estimate \cite{gej}.
Within the 'valence' scenario, on the contrary, a  negative light sea
helicity $\Delta \bar{u}=\Delta \bar{d} \equiv \Delta \bar{q}
\simeq -0.07$ alone suffices and we shall assume a maximally
$SU(3)_f$ broken polarized strange sea input
$\delta s(x,\mu^2)=\delta \bar{s}(x,\mu^2)=0$ here \cite{grsv}. 
Finally, we note that in both above scenarios the Bjorken sum rule 
manifestly holds due to the constraints (\ref{eq:stdsums1}),
(\ref{eq:valsums1}).
%
%
\begin{figure}[ht]
\begin{center}
\vspace*{-1.5cm}
\epsfig{file=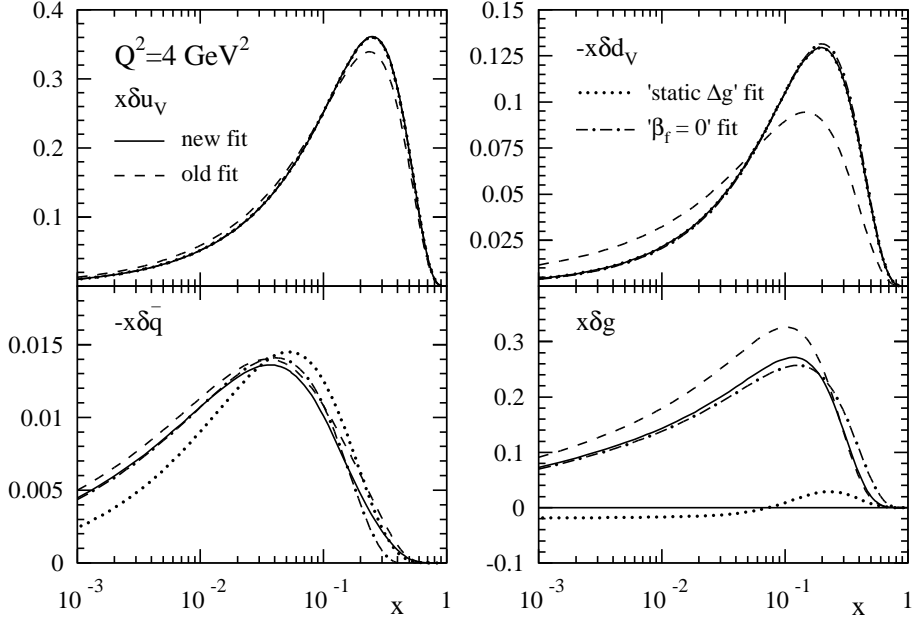, width=13.5cm}
\vspace*{-0.5cm}
\caption{\sf The polarized NLO $\overline{\mathrm{MS}}$ densities at 
$Q^2=4\,\mathrm{GeV}^2$ in the 'standard' scenario as obtained in our new 
and old \cite{grsv} analyses. Also shown are the distributions obtained
in two other fits employing some extra constraints on the input
distributions. It should be noted that as a result of the fits the
sea always turns out to be $SU(3)_f$ symmetric, i.e., 
$\bar{q}\equiv\bar{u}=\bar{d}=\bar{s}$.}
\vspace*{-0.5cm}
\end{center}
\end{figure}

%
%
\section{Quantitative NLO Results}
\noindent
Turning to the determination of the polarized NLO parton
distributions $\delta f(x,Q^2)$ it is helpful to consider some
reasonable and simple, but still sufficiently flexible ansatz for
the $\delta f(x,\mu^2)$ with not too many free parameters. As a 
general ansatz we take \cite{grsv} 
\begin{equation}
\label{eq:ansatz}
\delta f(x,\mu^2) = N_f x^{\alpha_f} (1-x)^{\beta_f} f(x,\mu^2)
\end{equation}
where $f=u_V, d_V, \bar{q}\equiv \bar{u}=\bar{d},s,g$ which
links the polarized input distributions to the unpolarized ones
as taken from \cite{grv94} (in this way the positivity 
requirements (\ref{eq:pos}) can be trivially implemented).
One can argue that the available inclusive DIS data do not allow for a
fully flavor-decomposed ansatz like (\ref{eq:ansatz}) but instead
only for a separation into non-singlet, singlet and gluon distributions.
This is of course true, but the aim of our analysis is not only to
simply fit the available data but also to provide a realistic set
of parton distributions which can be further probed in processes 
other than DIS, i.e., to have 'predictive power'.
For obvious reasons, we have not taken into account any 
$SU(2)_f$ breaking input $(\delta\bar{u}\neq\delta\bar{d})$ in
(\ref{eq:ansatz}).
Moreover the number of parameters in (\ref{eq:ansatz}) can be further 
reduced without any worsening of the fit by setting 
$\beta_{u_{V}}=\beta_{d_{V}}=0$, $\alpha_s=0$, and $\beta_s=0$ 
in (\ref{eq:ansatz}).
%
%
\begin{figure}[th]
\begin{center}
\vspace*{-1.9cm}
\epsfig{file=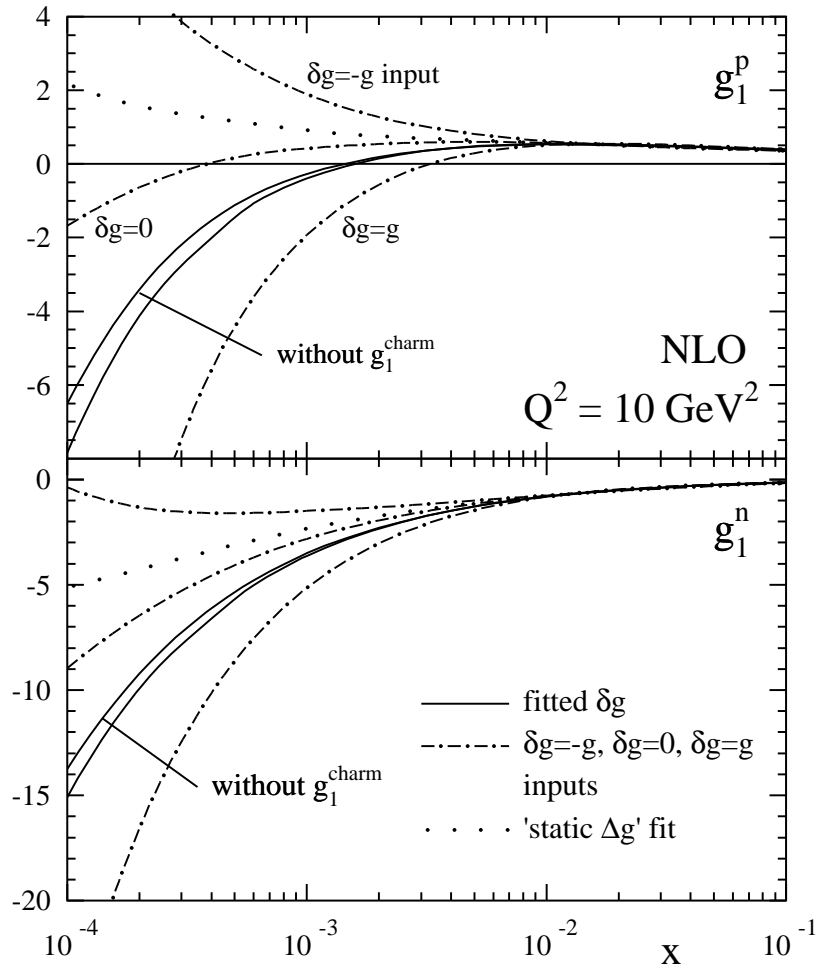, width=12.5cm}
\vspace*{-0.8cm}
\caption{\sf The small-$x$ behaviour of $g_1^p$ (upper part)
and $g_1^n$ (lower part) in NLO at $Q^2=10\,\mathrm{GeV}^2$ 
as predicted by various 'standard'
scenario fits employing different boundary conditions for the 
polarized input gluon distribution $\delta g(x,\mu^2)$. 
For the optimal fit input also the effect of not including the
charm contribution to $g_1^{p,n}$ is shown.}
\vspace*{-0.8cm}
\end{center}
\end{figure}

A comparison of our best fit employing the 'standard' scenario 
constraints (\ref{eq:stdsums1}) and (\ref{eq:stdsums2}) with the 
available data on $A_1^N(x,Q^2)$ [1-3,5,16] is presented in fig.\ 1. 
Also shown are the results for $A_1^N$ obtained by using our 
previous fit results \cite{grsv}. The results in the 'valence' scenario are 
indistinguishable from the ones shown and hence suppressed. 
The total $\chi^2$ of the 'new' and 'old' optimal 'standard' scenario fits is
123.02 and 144.37, respectively, for 168 data points. As can be
inferred from fig.\ 1 the results for the proton $(A_1^p)$ as well
as for the deuteron $(A_1^d)$ are basically not affected by the
reanalysis despite of the rather large differences in the neutron
case $(A_1^n)$ mainly due to the rather precise, new E154 data
\cite{e154}. 
This can be understood better by comparing the individual parton
distributions $\delta f(x,Q^2)$ rather than $A_1^N$ itself. This is
done in fig.\ 2 for the 'new' and 'old' \cite{grsv} NLO 
$\overline{\mathrm{MS}}$ densities at $Q^2=4\,\mathrm{GeV}^2$.
As can be seen from fig.\ 2 apart from a slightly smaller
gluon distribution $\delta g$ only $\delta d_V$ changes by quite a
large amount. This is because $\delta d_V$ is mainly probed in
$g_1^n$ (i.e., $A_1^n$) where we have observed the largest changes
as compared to our old results \cite{grsv} in fig.\ 1.

Also shown in fig.\ 2 are the results of two other fits which are based
on some additional constraints on the input distributions. 
For the '$\beta_f=0$' fit we have set
$\beta_f=0$ in our ansatz (\ref{eq:ansatz}). The total $\chi^2$ of 
123.6 is very similar to our best fit and this confirms the 
results of a recent fit by the E154 collaboration \cite{e154fit} based on our 
radiative parton model framework \cite{grsv} where all $\beta$'s where
chosen to be zero from the very beginning. The 'static $\Delta g$' fit
also yields a similar total $\chi^2$ (124.28) and the idea behind
this fit deserves a further explanation. 
If one studies the $Q^2$-evolution of the total gluon polarization
$\Delta g(Q^2)$ one observes that $\Delta g$ rises for increasing
values of $Q^2$ if one starts with an input
$\Delta g(\mu^2)>\Delta g_{static}$. On the other hand $\Delta g$
decreases for increasing $Q^2$ if $\Delta g(\mu^2)<\Delta g_{static}$ hence
remaining constant for {\em{all}} values of $Q^2$ for 
$\Delta g(\mu^2)=\Delta g_{static}$ (cf.\ fig.\ 5). In LO the precise value of
$\Delta g_{static}$ can be easily obtained from the evolution equation
for $\Delta g$ by setting $d\Delta g(Q^2)/d\ln Q^2 =0$. This yields
\begin{equation}
\label{eq:staticg}
\Delta g_{static} \simeq - \frac{1}{2} \Delta \Sigma \approx {\cal{O}}(-0.15)
\end{equation} 
where $\Delta \Sigma = \sum_q (\Delta q + \Delta \bar{q})$ is the
total helicity carried by quarks and antiquarks and 
(\ref{eq:staticg}) is only subject to a small correction in NLO. 
%
%
\begin{figure}[ht]
\begin{center}
\vspace*{-1.5cm}
\epsfig{file=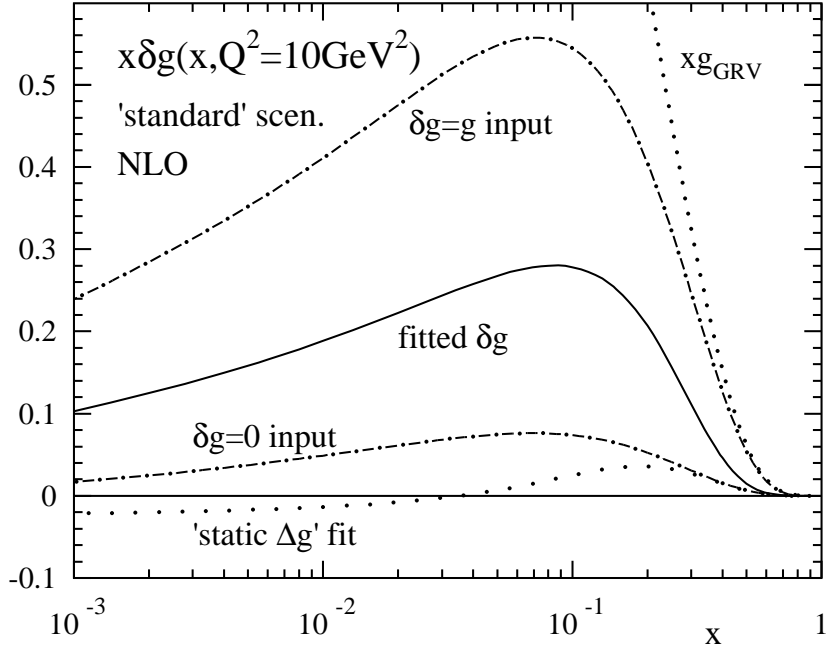, width=13cm}
\vspace*{-0.6cm}
\caption{\sf The experimentally allowed range of NLO polarized gluon
densities at $Q^2=10\,\mathrm{GeV}^2$ for the 'standard' scenario.
Also shown is the unpolarized gluon distribution of ref.\ \cite{grv94}.}
\vspace*{-0.5cm}
\end{center}
\end{figure}

Fig.\ 2 also reveals that the valence (and to some extent also the
sea) distributions are quite well determined by present data in our
new fits despite of the different underlying constraints for the
input distributions. On the contrary the polarized gluon density $\delta g$ is
still hardly constrained at all, even the rather small and
'exotic' $\delta g$ resulting from the 'static $\Delta g$' fit is not
excluded.

Inevitably the large uncertainty in $\delta g$ implies that
also the small-$x$ behaviour of $g_1$ beyond the experimentally 
accessible $x$-range is completely uncertain and not
predictable as is demonstrated in fig.\ 3 for the proton and the neutron case
in the 'standard' scenario for $Q^2=10\,\mathrm{GeV}^2$.
Apart from the optimal input and the already discussed 'static $\Delta g$' fit
we also present in fig.\ 3 the results obtained by chosing 
three other extreme boundary
conditions for $\delta g(x,\mu^2)$. The total $\chi^2$ values for these
'$\delta g=-g$', '$\delta g=0$', and '$\delta g=g$' inputs are
134.68, 124.24, and 127.44, respectively. Even these inputs give still
excellent fits and only the largest possible negative input
in the radiative parton model (cf.\ eq.(\ref{eq:pos})), i.e.,
$\delta g(x,\mu^2)=-g(x,\mu^2)$, is disfavored by its $\chi^2$ value.
As can be seen, all these extreme inputs give 
indistinguishable results for $g_1^{p,n}$ in the experimentally
covered $x$-region $(x\gtrsim 0.01)$ whereas they lead to a rather
large spread in the small-$x$ region making any predictions impossible
here. Note that one obtains completely similar results also in the 
'valence scenario'.
This uncertainty implies also a large theoretical error from
the extrapolation $x\rightarrow 0$ when calculating the first moments
$\Gamma_1^{p,n}$. A conservative theoretical estimate for $\Gamma_1^{p,n}$,
taking into account the maximally allowed spread 
in $g_1^{p,n}$ by the positivity requirement (\ref{eq:pos})
in the radiative parton model,
i.e., $'\delta g=-g' \ldots '\delta g=g'$ inputs, yields
\begin{equation}
\label{eq:smallx}
\Gamma_1^p(Q^2=10\,{\mathrm{GeV}}^2)=0.133\pm 0.008\;\;,\;\;
\Gamma_1^n(Q^2=10\,{\mathrm{GeV}}^2)=-0.062\pm 0.008 \;\,.
\end{equation}

Also shown in fig.\ 3 is the effect of not including the
charm contribution to $g_1^{p,n}$ for our optimal fit results. As
already mentioned in section 2, $g_1^{charm}$ as calculated
from the appropriate massive $\gamma^* g\rightarrow c\bar{c}$
subprocess \cite{watson} is negligibly small in the experimentally covered 
$x$-region and remains small even for lower values of $x$.

%
%
\begin{figure}[ht]
\begin{center}
\vspace*{-1.8cm}
\epsfig{file=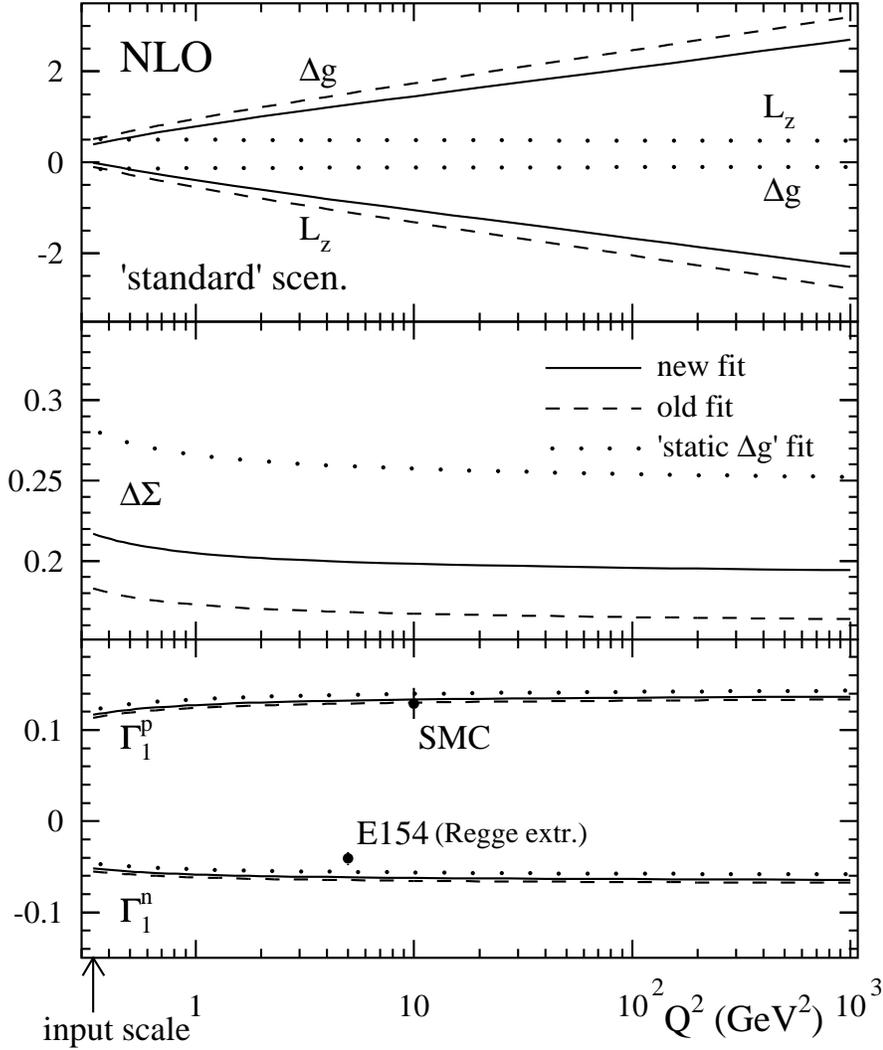,width=13.5cm}
\vspace*{-0.5cm}
\caption{\sf NLO $Q^2$-evolution of the first moments $\Delta g$,
$\Delta \Sigma$, and $\Gamma_1^{p,n}$ as well as of the
total orbital angular momentum contribution $L_z$ to the
helicity sum rule (\ref{eq:helisum}) from our low input scale
$\mu^2$ up to $Q^2=1000\,\mathrm{GeV}^2$ for our 'old' \cite{grsv}
and 'new' optimal fits as well as for the 'static $\Delta g$' input.
Also shown are some recent experimental results for $\Gamma_1^p$ and
$\Gamma_1^n$ from SMC \cite{smcnew} and E154 \cite{e154}, respectively.}
\vspace*{-0.5cm}
\end{center}
\end{figure}
Fig.\ 4 compares the different NLO gluon distributions 
at $Q^2=10\,\mathrm{GeV}^2$ as obtained from the various inputs 
discussed above. It is immediately obvious that other 
measurements apart from DIS are required to further pin down $\delta g$. 
For completeness
the total polarizations $\Delta g(Q^2)$ at $Q^2=10\,\mathrm{GeV}^2$ for 
the '$\delta g=g$', best fit $\delta g$, '$\delta g=0$', and
'static $\Delta g$' inputs are 3.2, 1.45, 0.31, and -0.12,
respectively. The value for the best fit gluon input is quite close
to the value of 1.74 found in our previous analysis \cite{grsv} and
is in agreement with other fits \cite{gs,abfr,e154fit}\footnote[2]{It
should be noted that the possibility of having a slightly negative
$\Delta g$ was recently also found in \cite{leader} but without
discussing the special 'static' properties of such a boundary
condition.}.

Finally, let us turn to the NLO $Q^2$-evolution of the first moments
$\Delta g(Q^2)$, $\Delta \Sigma(Q^2)$, and $\Gamma_1^{p,n}(Q^2)$ which is
shown for the 'standard' scenario in fig.\ 5 for $Q^2$ values ranging
from our low input scale $\mu^2$ up to $Q^2=1000\,\mathrm{GeV}^2$. Also
shown is the $Q^2$-dependence of the total orbital angular momentum
$L_z$ as obtained from the helicity sum rule
\begin{equation}
\label{eq:helisum}
\frac{1}{2}=\frac{1}{2}\Delta\Sigma(Q^2) + \Delta g(Q^2) +
L_z(Q^2)\;\;.
\end{equation}
It is interesting to observe that at our low input scale for our optimal
fit (this hold true also for our previous results \cite{grsv}) the
nucleon's spin is dominantly carried just by the total helicities of quarks
and gluons, i.e., $L_z(\mu^2)\approx 0$, and only during the 
$Q^2$-evolution a large negative $L_z(Q^2)$ is being built up in order
to compensate for the strong rise of $\Delta g(Q^2)$ in (\ref{eq:helisum}),
see fig.\ 5.

For our 'static $\Delta g$' fit the situation is completely different.
First of all one should note that $\Delta g(Q^2)$ is indeed independent
of $Q^2$ and quite small (cf.\ eq.\ (\ref{eq:staticg})) and this in turn
implies that also $L_z$ is practically $Q^2$-independent because 
$\Delta \Sigma(Q^2)$ is only weakly $Q^2$-dependent in the 
$\overline{\mathrm{MS}}$ scheme (it should be recalled that in LO
$\Delta \Sigma (Q^2)=\mathrm{const.}$). But more striking is the fact
that for this boundary condition the quark and gluon contributions
to the helicity sum rule (\ref{eq:helisum}) cancel each other
(see eq.\ (\ref{eq:staticg})) implying that for
{\em{all}} values of $Q^2$ the proton spin is entirely of 
angular momentum origin. This result is quite puzzling and completely
different from the intuitively expected vanishing of $L_z$ at some 
low bound-state-like scale as observed for our optimal fit but cannot
be excluded yet by the presently available fixed target data. 

%
\section{Summary and Outlook}
%
\noindent
We have presented an updated NLO QCD analysis of the DIS spin asymmetry
$A_1^N(x,Q^2)$ data in the $\overline{\mathrm{MS}}$ scheme in the
framework of the radiative parton model. Compared to our previous
results \cite{grsv} we have observed a rather large change in the 
polarized $\delta d_V$ distribution which is mainly due to new data
for $A_1^n$, in particular from E154 \cite{e154}. In contrast to the
polarized valence (and sea) quarks the gluon density $\delta g(x,Q^2)$
turns out to be still hardly constrained at all by present data. Our
optimal fits, however, still favor a rather sizeable total gluon
helicity, e.g., $\Delta g(Q^2=10\,\mathrm{GeV}^2)\simeq 1.45$, but it
was shown that even rather exotic boundary conditions for 
$\delta g$, with a rather small total helicity, such as the
'static $\Delta g$' input, yield excellent descriptions of all 
available $A_1^N(x,Q^2)$ data.

The latter gluon input has the striking consequence that the spin of
the nucleon is entirely made of orbital angular momentum for {\em{all}} 
values of $Q^2$ contrary to the intuitively expected vanishing of
$L_z$ at some low scale $\mu^2$ as observed for our optimal fits.
Finally, it was shown that the uncertainty in $\delta g$ induces a rather
large spread in the small-$x$ behaviour of $g_1^N$ making any 
predictions impossible. An estimate for the theoretical error
in the determination of $\Gamma_1^{p,n}$ due to the $x\rightarrow 0$
extrapolation uncertainty was given.

Future fixed target DIS data, in particular from E155, as well as 
semi-inclusive measurements from SMC and HERMES will help to pin
down the polarized quark distributions more precisely but it
cannot be expected that $\delta g$ can be further constrained by such
measurements since the lever-arm in $x$ and $Q^2$ is too limited for an
indirect determination of $\delta g$ from scaling-violations.
A realization of the currently discussed upgrade of HERA to a
polarized $ep$ collider would be extremely helpful in this
respect to gain more insight into $\delta g$ as well as to
pin down the small-$x$ behaviour of $g_1^p(x,Q^2)$ more precisely.
%
\section*{Acknowledgements}
\noindent
It is a pleasure to thank M.\ Gl\"{u}ck, E.\ Reya, and W.\ Vogelsang
for a fruitful collaboration. I am grateful to the University of
Dortmund where most of the work presented here was done. This work
was supported in part by the 'Bundesministerium f\"{u}r
Bildung, Wissenschaft, Forschung und Technologie', Bonn. 


%
\end{document}